\documentclass[showpacs,twocolumn]{revtex4}

\usepackage{graphicx}
\usepackage{dcolumn}
\usepackage{amsmath}

\makeatletter
\def\btt#1{\texttt{\@backslashchar#1}}
\DeclareRobustCommand\bblash{\btt{\@backslashchar}} \makeatother

\begin{document}

\title[]{Radiating Kerr-Newman black hole in  $f(R)$ gravity}
\author{Sushant~G.~Ghosh$^{a,\;b\;}$} \email{sghosh2@jmi.ac.in,
sgghosh@gmail.com}
\author{Sunil~D.~Maharaj$^{a}$}\email{maharaj@ukzn.ac.za}
\author{Uma Papnoi$^{b}$} \email{uma.papnoi@gmail.com} \affiliation{$^{a}$ Astrophysics and Cosmology
Research Unit, School of Mathematics, Statistics and Computer Science, University of
KwaZulu-Natal, Private Bag 54001, Durban 4000, South Africa}
\affiliation{$^{b}$ Centre for Theoretical Physics, Jamia Millia
Islamia, New Delhi 110025, India}
\date{\today}

\begin{abstract}
We derive an exact radiating  Kerr-Newman like black hole solution,
with constant curvature $R=R_0$ imposed, to {\it metric} $f(R)$ gravity
via complex transformations suggested by Newman-Janis. This
generates a geometry which is precisely that of radiating
Kerr-Newman-de Sitter / anti-de Sitter with the $f(R)$ gravity
contributing an $R_0$ cosmological-like term. The structure of three
horizon-like surfaces, {\it viz.} timelike limit surface, apparent horizon
and event horizon, are determined.  We demonstrate the existence of
an additional cosmological horizon, in $f(R)$ gravity model, apart
from the regular black hole horizons that exist in the analogous general
relativity case. In particular, the known stationary Kerr-Newman
black hole solutions of $f(R)$ gravity and general relativity are
retrieved.  We find that the timelike limit surface becomes less
prolate with $R_0$ thereby affecting the shape of the corresponding ergosphere.

\end{abstract}



\maketitle

\section{Introduction}
 The $f(R)$ gravity, where $f(R)$ is an analytic function of the Ricci
scalar $R$, comes into existence as a straightforward extension of
general relativity (GR) \cite{frr,frr1,scml}. In these theories the
curvature scalar $R$ of the Lagrangian in the Einstein-Hilbert
action is replaced by an arbitrary function of the curvature scalar
thereby modifying GR. However, the $f(R)$ action is sufficiently
general to encapsulate some of the basic characteristics of higher
order gravity. It is an interesting and relatively simple
alternative to GR,  the study of which yields useful conclusions
\cite{frr,frr1,scml}, including the present day acceleration. Unlike
GR, which demands metric function derivatives no higher than second
order, the {\it metric} $f(R)$ gravity has up to fourth order
derivatives \cite{lc}. This causes complication in the calculations
and hence, in general, finding exact solutions in this theory is
laborious.  Accordingly, little is known about $f(R)$ gravity exact
solutions, which deserve to be understood better. Nevertheless,
recently, interesting measures have been taken to get the
spherically symmetric solutions of $f(R)$ gravity
\cite{mv,cen,cdm,sz,pbon,asav,Moon,ncgd,cst08}. In particular,
spherically symmetric black hole (BH) solutions were obtained for a
positive constant curvature scalar in \cite{cen}, and a BH solution
was obtained in $f(R)$ gravities by requiring the presence of a
negative constant curvature scalar \cite{cdm}.  Also, several
spherical $f(R)$ BH solutions have been obtained
\cite{cdm,sz,pbon,asav,Moon}. The generalization of these stationary
$f(R)$ BHs to the axially symmetric case, Kerr-Newman BH, was
addressed recently \cite{cls,cdr,al,adds}. In particular, it is
demonstrated \cite{cls} that the rotating BH solutions for $f(R)$
gravity can be derived starting from exact spherically symmetric
solutions by a complex coordinate transformation previously
developed by Newman and Janis \cite{nj} in GR. However, the axially
symmetric $f(R)$ is still unexplored, e.g., the radiating
generalization of the $f(R)$ Kerr-Newman BH is still unknown. It is
the purpose of this paper to generate this metric and we also
present the $D$-dimensional Kerr metric in $f(R)$ gravity. Thus we
extend a recent work of ours \cite{sgsd} on radiating $f(R)$ BH to
include rotation. The Kerr metric \cite{kerr} is undoubtedly the
single most significant exact solution in the Einstein theory of GR,
which represents the prototypical BH that can arise from
gravitational collapse. The Kerr-Newman spacetime is associated with
the exterior geometry of a rotating massive and charged BH
\cite{ggsh}.  It is well known that Kerr BH enjoys many interesting
properties distinct from its nonspinning counterpart, i.e., from
Schwarzschild BH. However, there is a surprising connection between
the two BHs of Einstein theory, and was analysed by Newman and Janis
\cite{nja}. They demonstrated that applying a complex coordinate
transformation, it was possible to construct both the Kerr and
Kerr-Newman solutions starting from the Schwarzschild metric and
Reissner Nordstr$\ddot{o}$m, respectively \cite{nja}. For  a review on the
Newman-Janis algorithm  see, e.g., \cite{d'Inverno}.

In this paper, the Newman-Janis algorithm is applied to spherically
symmetric radiating $f(R)$ BH solutions, and the corresponding
radiating rotating solutions, namely radiating $f(R)$ Kerr-Newman
metrics  are obtained in Section III. We investigate further the
structure and location of horizons of the radiating $f(R)$
Kerr-Newman metric in Section IV. We consider whether $f(R)$ gravity
plays any special role in the formation of horizons in Section V.
 The general remarks on $f(R)$ gravity is given in Section II, and here we
also discuss briefly basic equations of HD $f(R)$ gravity.  The
paper ends with concluding remarks in Section VI.  We also give the
HD Kerr like metric, for constant curvature imposed, $f(R)$ gravity, in
the appendix. We use units which fix the speed of light and the
gravitational constant via $G = c = 1$, and use the metric signature
($+,\;-,\;-,\;-$).

\section{Basic Equations of \emph{Metric} $f(R)$ gravity}
In this section we briefly review the \emph{metric} $f(R)$ gravity
in higher-dimensional (HD) spacetime. The starting point is  the
modified Einstein-Hilbert, $D$-dimensional, gravitational action
\cite{cdm}:
\begin{equation}
I=\frac{1}{16 \pi}\int {d}^{D}x\sqrt{ - g }\,(R+f(R)),
 \label{action}
\end{equation}
where $g$ is the determinant of the metric $g_{ab}$, $(a,b=0, 1,
..., D-1)$, $R$ is the scalar curvature, and $f(R)$ is the real
function defining the theory under consideration.  As the simplest
example, the  Einstein-Hilbert action with cosmological constant
$\Lambda_f$ is given by $f(R)=-(D-2)\Lambda_f$.

From the above action, the equations of motion in the metric
formalism are just \cite{scml}:
\begin{eqnarray}
& & R_{ab}(1+f'(R)) - \frac{1}{2}(R+f(R))\,g_{ab} \nonumber \\ + &&
( g_{ab} \nabla^2 - \nabla_a \nabla_b  )f'(R) = 2 T_{ab},
\label{EEOM}
\end{eqnarray}
where $R_{ab}$ is the usual Ricci tensor, the prime in $f'(R)$
denotes differentiation  with respect to $R$, and $\nabla^2
=\nabla_a \nabla^a$ with $\nabla$ being the usual covariant derivative.
 Here, we are
interested in obtaining the constant scalar curvature solutions
$R\,=\,R_0$. Taking the trace in the Eq.~(\ref{EEOM}), we get
\begin{eqnarray}
 2(1+f'(R_0))\,R_{0}-D\,(R_{0}+f(R_{0}))\,=\,0,
\label{ER0}
\end{eqnarray}
where we assume that $T=T^a_a=0$ and note that $g_a^a=\delta^a_a=D$.
Eq.~ (\ref{ER0}) determines the negative constant curvature scalar
as \cite{cdm}
\begin{equation}
R_0=\frac{D\,f(R_0)}{2(1+f'(R_0))-D} \equiv D \Lambda_f.
\label{const}
\end{equation}
Thus any constant curvature solution $R=R_0$ with $1+f'(R_0)\neq 0$
obeys \cite{scml}
\begin{equation}\label{EEOM1}
R_{ab} = \Lambda_f g_{ab} + \frac{2}{1+f'(R_0)} T_{ab}.
\end{equation}
For this kind of solution an effective cosmological constant may be
defined as $\Lambda_f\equiv R_{0}/D$ for $D$-dimensional spacetime.
In this paper we consider the case with conformal matter
($T=T_a^a=0$). For the case of conformal matter with non-vanishing
$\Lambda_f$ we have again constant $R=R_0$ with $R_0=D \Lambda_f$
and $g_{ab}$ is a solution of $f(R)$ provided that once again
$f(D\Lambda_D)=\Lambda_D(2-D+2f'D\Lambda_D)$.  In this case the
solution is dS or (A)dS depending on the sign of $R_0$, just as in
GR with a cosmological constant.

We note that the condition $1 + f'(R_0) > 0$ is required to avoid the
appearance of ghosts \cite{frr},  and a necessary condition that the
stationary $f(R)$ BH becomes a type of Schwarzschild-(A)dS BH.
Further, we require that $f''(R) > 0$ to avoid the negative mass
squared of a scalar-field degree of freedom, i.e., to avoid a
tachyonic instability \cite{frr,lc}. Also $f(R)$ is a monotonic
increasing function with $-1 < f'(R) < 0.$

\section{Radiating Rotating $f(R)$ BH via Newman-Janis}
Let us begin with the action for $f(R)$ gravity with a Maxwell term
in the 4D case \cite{Moon}:
\begin{eqnarray} \label{ActionEm}
I_g=  \frac{1}{16\pi }\int d^4 x\sqrt{-g} \Big[ R+f(R)-F_{ab}F^{ab}
\Big].
\end{eqnarray}
The Maxwell tensor is $F_{ab} = \partial_a A_b - \partial_b A_a$,
where $A_a$ is the vector potential. From the variation of the above
action (\ref{ActionEm}), the Einstein equation of motion can be
written as \cite{cdm,Moon}
\begin{eqnarray} \label{equaem}
R_{ab}\Big(1+f'(R)\Big)-\frac{1}{2}\Big(R+f(R)\Big)g_{ab} \nonumber
\\ + \Big(g_{ab}\nabla^2-\nabla_{a}\nabla_{b}\Big)f'(R)=2
T_{ab}^{EM},
\end{eqnarray}
with the EMT for charged null dust \cite{dg}
\begin{equation}\label{emt}
T_{ab}^{EM}=\zeta(v,r) n_{a}n_{b} +
 F_{a\rho}F_{b}~^{\rho}-\frac{g_{ab}}{4}F_{\rho\sigma}F^{\rho\sigma}.
\end{equation}
Once again, it is easy to see that the trace of the EMT is
$T^{EM}=0$, due to the fact that $F^a_a=0$ in 4D.  However, in HD
$T^{EM}\neq 0$. On the other hand, the Maxwell equations take the
form $ \nabla_{a}F^{ab}=0.$ To proceed further, with constant
curvature constant $R_0$, and  taking the trace of the
Eq.~(\ref{equaem}),  after some algebra, leads to
\begin{equation}\label{roem}
R_0 = \frac{2 f(R_0)}{f'(R_0) - 1} \equiv 4 \Lambda_f.
\end{equation}
Thus any constant curvature solution $R=R_0$ obeys
\begin{equation}\label{EEOM2}
R_{ab} = \Lambda_f g_{ab} + \frac{2}{1+f'(R_0)} T_{ab},
\end{equation}
and an effective cosmological constant may be defined as $\Lambda_f
\equiv R_{0}/4$.

Here we wish to obtain general $f(R)$ radiating rotating  BH
solution from a spherically symmetric BH solution via the complex
transformation suggested by Newman-Janis \cite{nja}. For this
purpose, We begin with  the ``seed metric", expressed in terms of the
Eddington (ingoing) coordinate $v$, as:
\begin{equation}
ds^{2} = e^{\psi(v,r)}dv\left[f(v,r)e^{\psi(v,r)}dv + 2  dr\right] -
r^2 d\Omega^2, \label{Metric}
\end{equation}
with $d\Omega^2 = d \theta^2+ \sin^2 \theta d \phi^2$. Here
$e^{\psi(v,r)}$ is an arbitrary function. It is useful to introduce
a local mass function $m(v,r)$ defined by $f(v,r) = 1 - {2
m(v,r)}/{r}$. For $m(v,r) = M(v)$ and $\psi(v,r)=0$, the metric
reduces to the standard Vaidya metric. We can always set without any
loss of generality, $\psi(v,r) = 0.$  Thus any spherically symmetric
radiating BH is defined by the metric (\ref{Metric}).  The function
$f(v,r)$ is a function of $v$ and $r$, and depends on the matter field
and the underlying theory and on functional form of $f(R)$ and  Ricci
curvature scalar $R=R(v,r)$.

The Newman-Janis algorithm can be applied to any spherically
symmetric static solution, generating rotating spacetimes \cite{nj}. For
example, The Kerr metric can be obtained from the Schwarzschild
metric, and Reissner-Nordstr$\ddot{o}$m solution leads to the
Kerr-Newman solutions \cite{nj,d'Inverno} which is based on a complex coordinate
transformation.  Recently, the axially symmetric model for $f(R)$-gravity is
derived from exact spherically symmetric $f(R)$-gravity solutions
\cite{cls,Mdl}. In the following, we apply the Newman-Janis
algorithm to the general spherically symmetric radiating BH given by
Eq. (\ref{Metric}) in order to construct a general radiating
rotating BH solution.

The metric $\tilde{g}_{ab}$ given by Eq. (\ref{Metric}) can be
written in terms of a null tetrad \cite{nja,cls} as:
\begin{equation}
\tilde{g}^{ab} = -L^a N^b - L^b N^a + M^a \bar{M}^b + M^b \bar{M}^a
\label{NPmetric}
\end{equation}
This tetrad is orthonormal obeying the conditions \begin{eqnarray}
&&L_a M^a = L_a \bar{M}^a = N_a M^a = N_a \bar{M}^a = 0, \\
&&L_a L^a = N_a N^a = M_a M^a = \bar{M}_a \bar{M}^a = 0, \\
&&L_a N^a = -1,~~~ M_a \bar{M}^a = 1. \end{eqnarray}
Next, we perform the similar complex coordinate transformation as
used by Newman and Janis \cite{nja}:
\begin{equation}\label{transf}
{x'}^{\mu} = x^{\mu} + ia (\delta_r^{\mu} - \delta_u^{\mu})
\cos\theta \rightarrow \\ \left\{\begin{array}{ll}
v' = v - ia\cos\theta, \\
r' = r + ia\cos\theta, \\
\theta' = \theta,~~~\phi' = \phi. \end{array}\right.
\end{equation}
and also transform the tetrad $Z^a_s = (L^a,\; N^a,\;M^a,\;
\bar{M}^a)$ in the usual way
\begin{equation} Z'^a_s =
\frac{\partial x'^a}{\partial x^b} Z^b_s,
\end{equation} which leads to
\begin{eqnarray}
L^a &=& \delta^a_r, \\
N^a &=&   \left[ \delta^a_v - \frac{1}{2} f(v,r) \delta^a_r \right], \\
M^a &=& \frac{1}{\sqrt{2}r} \left[ ia\sin\theta \left(\delta^a_v - \delta^a_r \right) + \delta^a_\theta + \frac{i}{\sin\theta} \delta^a_\phi \right], \\
\bar{M}^a &=& \frac{1}{\sqrt{2}\bar{r}} \left[ -ia\sin\theta
\left(\delta^a_v - \delta^a_r \right) + \delta^a_\theta -
\frac{i}{\sin\theta} \delta^a_\phi \right], \end{eqnarray}
 and, we have dropped the primes.
 This transformed tetrad yields a new metric given by the line
 element (see Ref. \cite{nja,cls}, for further details):
\begin{eqnarray}\label{rotbh}
ds^2 & = & \mathcal{F}(v,r,\theta) dv^2 - \Sigma(r,\theta) d
\theta^2 + 2\;[dv-  a \sin^2 \theta d\phi]\;dr  \nonumber \\ & - & \left[ a^2
(2- \mathcal{ F}(v,r,\theta) )\sin^2 \theta  + \Sigma(r,\theta)
\right] \sin^2 \theta d\phi^2 \nonumber \\ & + & 2 a \left[ 1 -
\mathcal{ F}(v,r,\theta) \right] \sin^2 \theta dv \; d\phi.
\end{eqnarray}
Here $\mathcal{F}(v,r,\theta) $ is function which depends on
$f(r,v)$, e.g., for the function $f(v,r) = 1 - \frac{2M(v)}{r} +
\frac{2  \tilde{Q}^2(v)}{r^2}$, it has the form $\mathcal{F}(v,r,\theta) =
1 - \frac{2M(v)r }{\Sigma } + \frac{2  \tilde{Q}^2(v)}{\Sigma }$,  where
$\Sigma \equiv r^2 + a^2 \cos^2\theta$.

The spherically symmetric collapsing solution of the $f(R)$ gravity
can be classified according to choice of the Ricci curvature
\cite{scml}.  The prototype of general pure $f(R)$ gravity
(nonconstant curvature) are the models  viz. $f(R)=R-\mu^4/R$,
$f(R)=R+\kappa R^n$, $f(R)=R- \lambda \exp(-\zeta R)$, $f(R)=R-
\lambda \exp(-\zeta R)+\kappa R^2$ and $f(R) = R-\lambda \exp(-\zeta
R)+\kappa R^n+ \eta \ln R$, where $\mu,\; \kappa,\; \lambda,\; \zeta
$ and $\eta$ are some constants \cite{scml}.  Hendi {\it et al.}
\cite{shh} obtained static solutions for these choices of $f(R)$ in
the pure $f(R)$-gravity and demonstrated that starting from pure
$f(R)$ gravity, with above mentioned choice of $f(R)$, leads to
(charged) Einstein-$\Lambda$ solutions which can be interpreted as
(charged) (a)dS BH solutions \cite{shh}. Following \cite{shh}, we
find the corresponding radiating solution \cite{up&sgg} and it turns
out that the solution is given by the metric ~(\ref{Metric}), with
\begin{equation}\label{emsol1}
f(v,r) = \kappa - \frac{2M(v)}{r} + \frac{2 \tilde{Q}^2(v)}{r^2} -
\frac{\Lambda}{3}r^2,
\end{equation}
where $M(v)$ and $\tilde{Q}(v)$ are functions of integration and $\Lambda$
may be interpreted as the cosmological constant emerging from $f(R)$
gravity and we choose  $\Lambda = R_0/4$ and also $\kappa=1$.  The function $\tilde{Q}(v)$ can
be zero for some choice of $f(R)$, i.e., for the model
$f(R)=R+\kappa R^n$, we cannot obtain charged solutions, i.e., in
this case the solution ~(\ref{emsol1}) is valid with $\tilde{Q}(v)=0$.
Thus it is clear that the general radiating spherically symmetric
solution in several $f(R)$ gravity models, with  the choice of
$f(R)$ mentioned above, result to the metric (\ref{Metric}) with
$f(v,r)$ as (\ref{emsol1}).

Now we start with the radiating spherically symmetric metric
(\ref{Metric}), written in Eddington-Finkelstein coordinate, with
$f(v,r)$ as given by (\ref{emsol1}).  By performing the Newman-Janis
algorithm, we derive the $f(R)$ radiating rotating solution which takes
the form:
\begin{eqnarray} \label{rknm}
ds^2&=& \frac{A}{\Sigma}\left[\Delta - \Theta a^2 \sin^2 \theta \right] dv^2+ 2 \sqrt{A} \left[ dv - a \sin^2 d\phi \right] dr  \nonumber  \\
   & & - \frac{\Sigma}{\Theta} d \theta^2 + A \frac{2 a}{\Sigma}\left[\Delta (r^2 + a^2)-\Theta \right]\sin^2\theta dv d\phi \nonumber \\
   & & - \frac{A}{\Sigma}\left[\Delta (r^2 + a^2)^2 -\Theta a^2 \sin^2\theta \right]
   \sin^2\theta
   d\phi^2,
\end{eqnarray}
where
\begin{eqnarray*}
\Sigma^2 &= & r^2+a^2\cos^2\theta,\\
\Delta &=&r^2+a^2-2M(v) r+\tilde{Q}^2(v)-\frac{R_0r^2}{12}(r^2+a^2),\\
\Theta &=&1+\frac{R_0}{12}a^2\cos^2\theta, \\
A &=&\left(1+\frac{R_0}{12}a^2\right)^{-2}, \\
\tilde{Q}^2(v) &=& \frac{2 {Q}^2(v)}{1+f'(R_0)},
\end{eqnarray*}
and the related electromagnetic potential is
\begin{equation}\label{em}
A_a = \frac{\tilde{Q}(v) r}{\Sigma^2} \left[\sqrt{A},\;0,\;0,\; -
\sqrt{A}a \sin^2\theta \right].
\end{equation}
Here $M(v)$ and $\tilde{Q}(v)$ are functions of retarded time $v$
identified, respectively, as mass and charge of spacetime, and $a$ is
the angular momentum per unit mass.  We have applied
the aforesaid procedure to a class of $f(R)$ gravity radiating
models. But, the method is general and is applicable to any general
radiating spherically symmetric solution in $f(R)$ gravity.  It
describes the exterior field of the radiating rotating charged body.
Thus we have a kind of charged radiating rotating metric in de Sitter/
anti-de Sitter (dS/AdS) like spacetime or radiating Kerr-Newman dS /
AdS like solution. The stationary $f(R)$ Kerr-Newman BH \cite{cdr,al,adds} in
$(t,\;r,\;\theta,\;\phi)$ can be obtained by means of the local
coordinate transformation and replacing $M(v)$ and $\tilde{Q}(v)$ by
constants $M$ and $Q$. The  metric (\ref{rknm}) of radiating $f(R)$
Kerr-Newman BH is a natural generalization of  the stationary
Kerr-Newman BH solutions of $f(R)$ gravity \cite{cdr,al}, but it is
Petrov type-II, whereas the latter is of Petrov type-D.  In addition, if
$R_0=0$ then metric~(\ref{rknm}) makes the Kerr-Newman metric. Hence, we
refer to the solution as a radiating $f(R)$ Kerr-Newman solution representing
gravitational collapse of a charged null fluid in a non-flat dS/AdS
like spacetime. Thus, the metric (\ref{rknm}) bears the same relation to
Kerr-Newman as does Vaidya metric to Schwarzschild metric. Also for
$Q=0$, the metric (\ref{rknm}) is radiating $f(R)$ Kerr spacetime
\cite{al,cdr}. If in addition $R_0 \rightarrow 0$, we have Kerr
spacetime and for $a \rightarrow 0$ the metric~(\ref{rknm}) is
$f(R)$ Bonnor-Vaidya spacetime which has zero angular momentum
\cite{sgsd}. The radiating rotating charged solution discussed here
is derived under the assumption of the constant curvature $R=R_0$,
in which case the $f(R)$ models are equivalent to GR and a
cosmological constant,  and also the solution is time-dependent.

\section{Singularity and physical parameters of radiating $f(R)$ Kerr-Newman BH}
The metric of the radiating $f(R)$ Kerr-Newman BH solution has the
form (\ref{rknm}) with electromagnetic  potential given by
(\ref{em}) and  the energy momentum tensor (\ref{emt}).  Here, we
shall discuss the singularity structure of radiating $f(R)$
Kerr-Newman BH derived in the previous section. The easiest way to
detect a singularity in a spacetime is to observe the divergence of
some invariants of the Riemann tensor.  We approach the singularity
problem by studying the behaviour of the Ricci $\mbox{R} = R_{ab}
R^{ab}$, ($R_{ab}$ the Ricci tensor) and Kretschmann invariants
$\mbox{K} = R_{abcd} R^{abcd}$, ($R_{abcd}$ the Riemann tensor). For
the metric (\ref{rknm}) they behave as:-
\begin{eqnarray}
\mbox{R}  \approx \frac{\mathcal{F}\left(\tilde{Q}(v),R_0\right)}{(r^2+a^2\cos^2\theta)^4}, \nonumber \\
\mbox{K} \approx  \frac{\mathcal{G}\left(M(v),\; \tilde{Q}(v),\;
a,\; \cos \theta,\; R_0 \right) }{(r^2+a^2\cos^2\theta)^6},
\label{eq:ks}
\end{eqnarray}
where $\mathcal{F}$ and $\mathcal{G}$ are some functions. It is
sufficient to study the Kretschmann and Ricci scalars for the
investigation of the spacetime curvature singularity(ies). These
invariants are regular everywhere except at the origin $r=0$ but
only at the equatorial plane $\theta=\pi/2$ for $a,\; M(v),\; $ and
$ \tilde{Q}(v) \neq0$. Hence, the spacetime has the scalar polynomial
singularity \cite{he} at $r=0$. The study of causal structure of the
spacetime is beyond the scope of this paper and will be discussed
elsewhere.

In order to further discuss the physical nature of radiating $f(R)$
Kerr-Newman BH, we introduce their kinematical parameters. Following
\cite{bc,jy,rm,bdk,xd98,xd99}, the null-tetrad of the metric
(\ref{rknm}) is of the form
\begin{eqnarray*} \label{nvector}
  l_a &=& \left[ \sqrt{A},\;0,\;0,\; - \sqrt{A} a \sin^2 \theta\right], \\
  n_a &=& \left[\sqrt{A} \frac{\Delta}{2 \Sigma},\;1,\;0,\; \sqrt{A} \frac{\Delta}{2 \Sigma}a \sin^2 \theta \right],\\
  m_a &=& \frac{\sigma}{\sqrt{2}\rho}\left[\sqrt{A} i a \sin \theta,\; 0,\; \frac{\Sigma}{\Theta},\; - \sqrt{A}i(r^2+ a^2) \sin \theta  \right],\\
  \bar{m}_a &=& \frac{\bar{\sigma}}{\sqrt{2}\bar{\rho}}\left[-\sqrt{A} i a \sin \theta,\; 0,\; \frac{\Sigma}{\Theta},\;  \sqrt{A}i(r^2+ a^2) \sin \theta
  \right],
\end{eqnarray*}
where
\begin{eqnarray*}
  \rho &=& r + i a \cos \theta, \\
  \sigma &=& 1 + i \sqrt{\left(\frac{R_0}{12}\right)} a \cos \theta,
\end{eqnarray*}
and $\bar{\rho}$ and $\bar{\sigma}$ are complex conjugates of,
respectively, $\rho$ and $\sigma$.  The null tetrad obeys null,
orthogonal and metric conditions
\begin{eqnarray}
l_{a}l^{a} & = & n_{a}n^{a} = m_{a} m^{a} = 0, \; ~l_a n^a = 1, \nonumber \\
l_{a}m^{a} & = & n_{a}m^{a} = 0, \; m_{a} \bar{m}^{a} = -1, \nonumber \\
g_{ab} & = & l_{a}n_{b} + l_{b}n_{a} -  m_{a} \bar{m}_{b} - m_{b}
\bar{m}_{a},\nonumber \\
g^{ab} & = & l^{a}n^{b} + l^{b}n^{a} -  m^{a} \bar{m}^{b} - m^{b}
\bar{m}^{a}.
\end{eqnarray}
Inspired by the arguments in  Ref.~\cite{bc,jy}, a null-vector
decomposition of the $f(R)$-metric (\ref{rknm}) is of the form
\begin{equation}\label{gab}
g_{ab} = - n_a l_b - l_a n_b + \gamma_{ab},
\end{equation}
where $\gamma_{ab} = m_{a} \bar{m}_{b} + m_{b} \bar{m}_{a}$. Next we
construct all physical parameters which help us to discuss the horizon
structure of radiating $f(R)$ Kerr-Newman BH. The optical behavior
of null geodesics congruences is governed by the Raychaudhuri
equation \cite{jy,rm,bdk,xd98,xd99}.
\begin{equation}\label{re}
   \frac{d \Theta}{d v} = \kappa \Theta - R_{ab}l^al^b-\frac{1}{2}
   \Theta^2 - \sigma_{ab} \sigma^{ab} + \omega_{ab}\omega^{ab},
\end{equation}
with expansion $\Theta$, twist $\omega$, shear $\sigma$, and surface
gravity $\kappa$. The expansion \cite{jy} of the null rays,
parameterized by $v$, is given by
\begin{equation}\label{theta}
\Theta = \nabla_a l^a - \kappa,
\end{equation}
where $\nabla$ is the covariant derivative. In the present case, the
surface gravity \cite{jy} is
\begin{equation}\label{sg}
\kappa = - n^a l^b \nabla_b l_a,
\end{equation}
and the shear \cite{jy} takes form
\begin{equation}\label{shear}
\sigma_{ab} =  \Theta_{ab} - \Theta (\gamma_c^c) \gamma_{ab}.
\end{equation}
The luminosity due to loss of mass is given by $L_M = - dM/dv$, $L_M
< 1$ , and due to gauge charge by $L_Q = - dQ/dv$, where $L_M, L_Q <
1$.  Both  are measured in the  region where $d/dv$ is timelike
\cite{jy,rm,bdk}.

\section{Three kinds of horizons}
If one considers $f(R)$ theory as modification of GR, it is natural
to discuss not only BH solutions but it's various properties in this
theory. It is expected that some features of BH may get modified in
$f(R)$ theories.  In this section we explore horizons
of radiating $f(R)$ Kerr-Newman BH, discuss the effects which come
from the $f(R)$ theories.  A BH has three horizon-like surfaces
\cite{jy}: timelike limit surface (TLS), apparent horizon (AH) and
event horizon (EH). For a classical Schwarzschild BH (which does
not radiate), the three surfaces EH, AH and TLS are all identical.
Upon "switching on" the Hawking evaporation this degeneracy is
partially lifted even if the spherical symmetry stays, e.g., for
Vaidya radiating BH, we have then AH=TLS,  but the EH is different
If we break spherical symmetry preserving stationarity (e.g., Kerr
BH), then AH=EH but EH $ \neq $ TLS.  In general, e.g., for
radiating Kerr-Newman BH, the three surfaces AH $\neq$ TLS $\neq$ EH
and they are sensitive to small perturbations.

 Here we are interested in  these horizons for the radiating $f(R)$ Kerr-Newman BH.
 As demonstrated first by York \cite{jy}, the horizons may be obtained to
 $O(L_M,L_Q)$ by noting that (i) for a BH with small dimensionless accretion,
 we can define  TLS's
 as locus where $g(\partial_v, \partial_v) = g_{vv} = 0$ (ii) AHs are defined as
surface such that $\Theta \simeq 0$ and (iii) EHs are surfaces such
that $d \Theta /dv \simeq 0$.

\begin{widetext}

\begin{figure}
\begin{tabular}{|c|c|c|}
\hline
\includegraphics[width=9 cm, height=6 cm]{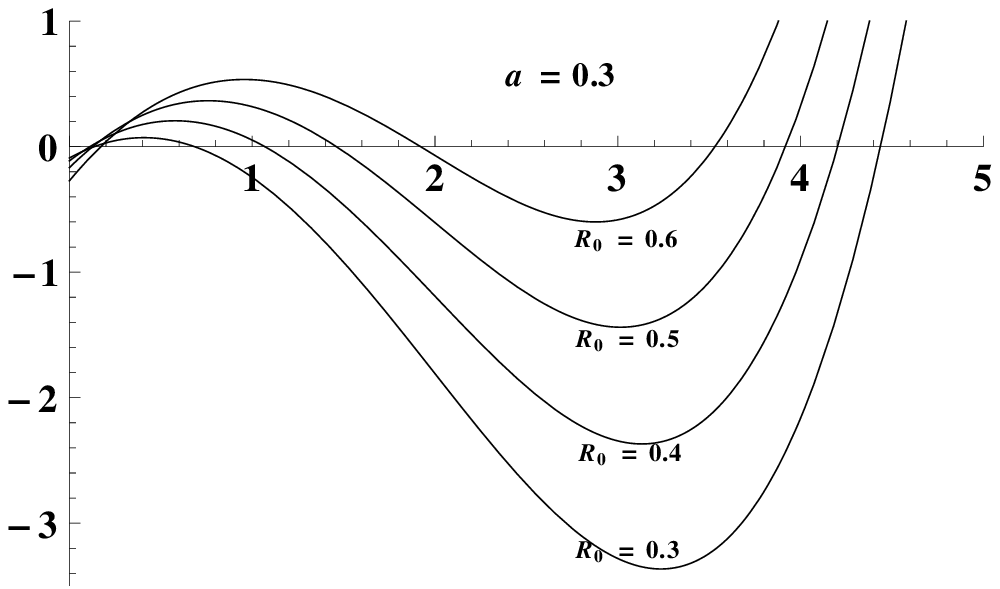}&
\includegraphics[width=9 cm, height=6 cm]{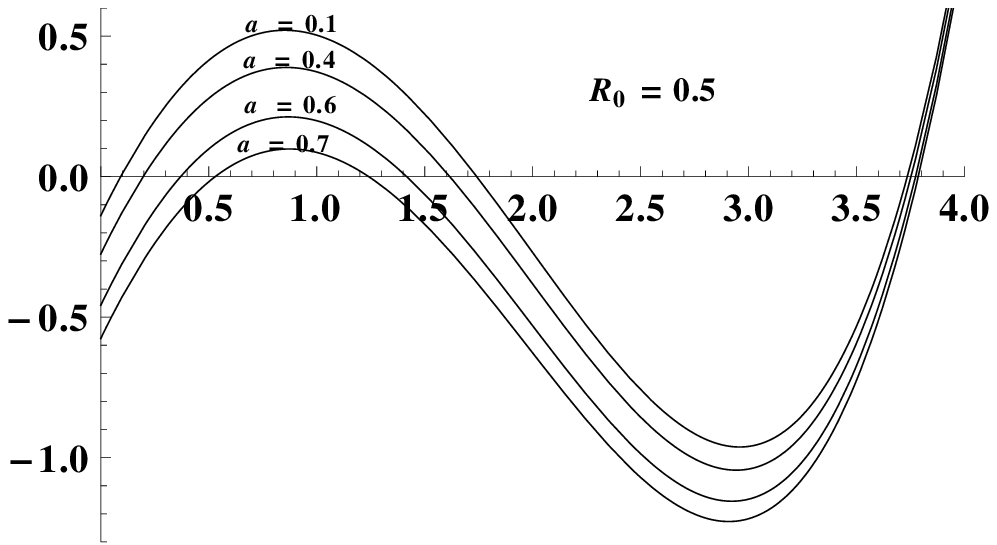}
\\
\hline
\end{tabular}
\caption{\label{TLS}TLS of radiating $f(R)$ Kerr-Newman BH: Plots showing positive
roots of Eq.~(\ref{tls1}) which corresponds TLS of radiating Kerr-Newman BH for $M(v)= \lambda v + O(v)$ and $\tilde{Q}(v)= \mu^2 v^2 + O(v^2)$ with
parameters values $\lambda= 0.04,\; \mu= 0.08$,\; and $\theta=0.3$. The three roots corresponds to $r_{TLS}^{-}$ (inner) and
$r_{TLS}^{+}$ (outer), and $r_{dTLS}$ of radiating $f(R)$ Kerr-Newman BH. LEFT: For different values of $R_0$ with $a=0.3$. RIGHT: For different values of $a$ with $R_0=0.5$.  }

\end{figure}

\end{widetext}

\begin{figure}
\includegraphics[width=10cm,angle=0]{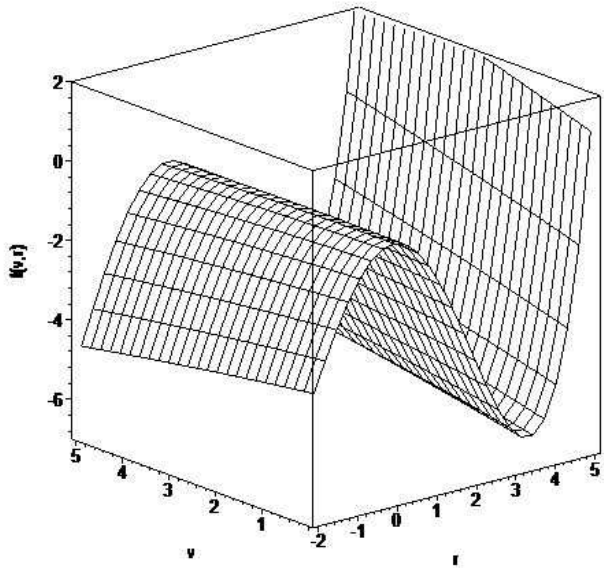}
\caption{A plot of the function $\mathfrak{f}(v,r,\theta)$ for
$M(v)= \lambda v + O(v)$ and $\tilde{Q}(v)= \mu^2 v^2 + O(v^2)$ with
parameters values $\lambda= 0.04,\; \mu= 0.08,\; \theta=0.3,\ a=0.7$
and $R_0=0.3$.  The zeros of $\mathfrak{f}(v,r,\theta) =0$
determines the TLS of radiating $f(R)$ Kerr-Newman BH} \label{TLSV}
\end{figure}

\subsection{Time-like limit surface}
The TLS is defined as the surface where
the static observer become light-like and it can be null, timelike or
spacelike \cite{jy}. First, we calculate the location of TLS surface,
which for the nonstationary radiating $f(R)$ Kerr-Newman metric
requires that prefactor of the $dv^2$ term in metric  vanishes; It
follows from Eq.~(\ref{rknm}) that TLS will satisfy \cite{xd99}
\begin{equation}\label{tls}
\Delta_r - \Delta_{\theta} a^2 \sin^2 \theta  = 0.
\end{equation}
This equation can be rewritten as
\begin{eqnarray}\label{tls1}
\mathfrak{f}(v,r,\theta)  = \frac{R_0}{12}r^4 - \left(1- \frac{R_0
a^2}{12} \right)r^2+ 2 M(v) r \nonumber
\\   + \frac{R_0}{12} \cos^2 \theta \sin^2\theta a^4 - \cos
\theta^2 a^2 - \tilde{Q}^2(v)=0,
\end{eqnarray}
Equation~(\ref{tls1}) is a reduced quartic equation.  It is easy to
check, under condition  of the discriminant in \cite{cond},
Eq.~(\ref{tls1}) admits four real roots.  For positive curvature
$R_0 > 0$, Eq.~(\ref{tls1}), subject to restriction \cite{cond}, has
all four real roots with three positive and one negative.   In
the Fig.~\ref{TLS}, we show three positive roots of the
Eq.~(\ref{tls1}).  The other three positive roots corresponds to
$r_{TLS}^{-}$ (inner) and $r_{TLS}^{+}$ (outer), and $r_{dTLS}$
(dS-like TLS). Clearly, $r_{TLS}^{-}\; <  \; r_{TLS}^{+}\; < \;
r_{dTLS}$ and that $r_{TLS}^{-}$ and $r_{TLS}^{+}$ are TLSs of a BH,
whereas the root $r_{dTLS}$ is supplementary TLS due to the $f(R)$
gravity term.  When $R_0 = 0$, i.e., in GR limit, have just two
outer and inner TLSs of radiating Kerr-Newman BH.

As mentioned above, in the limit $a \rightarrow 0$, one gets $f(R)$
Bonnor-Vaidya solution \cite{sgsd}, and Eq.~(\ref{tls1}) takes the
form
\begin{equation}\label{tlsnr}
\frac{R_0}{12}r^4 - r^2 + 2 M(v)r - \tilde{Q}^2(v)=0.
\end{equation}
This coincides with the nonrotational case in which case the various
horizons are identified and analyzed by us in \cite{sgsd} and hence,
to conserve space, we shall avoid the repetition of same. Further,
in the GR limit $R_0 \rightarrow 0; \tilde{Q}(v)\rightarrow Q(v) $,
we obtain
\begin{equation}\label{gr}
{r}^{2}+ \cos ^{2}\left( \theta \right) {a}^{2}-2\,M
 \left( v \right) r+ Q^{2} \left( v \right)  =0,
\end{equation}
which trivially solves to
\begin{eqnarray} \label{tlskn}
  r_{TLS}^{-} &=& M (v) -\sqrt {  M^{2}(v)   -
 {a}^{2} \cos^{2} \theta   - Q^{2}(v)},
 \nonumber \\
  r_{TLS}^{+} &=& M (v) + \sqrt {M^{2}(v)  -
 {a}^{2} \cos^{2} \theta   - Q^{2}(v)}.
\end{eqnarray}
These are regular outer and inner TLSs for a radiating Kerr-Newman
BH \cite{ggsh}, and further  in the non-rotating limit $a
\rightarrow 0$, the solutions (\ref{tlskn}) reduces to
\begin{eqnarray} \label{tlskn1}
  r_{TLS}^{-} &=& M (v) -\sqrt {  M^{2}(v)   -
   Q^{2}(v)},
 \nonumber \\
  r_{TLS}^{+} &=& M (v) + \sqrt {M^{2}(v)    - Q^{2}(v)},
\end{eqnarray}
which are TLS of Bonnor-Vaidya BH.  Thus the radiating $f(R)$
Kerr-Newman BH, in the GR limit and $a \rightarrow 0$, degenerates
to Bonnor-Vaidya BH \cite{dg}.

\begin{widetext}

\begin{figure}
\begin{tabular}{|c|c|}
\hline
\includegraphics[width=9 cm, height=6 cm]{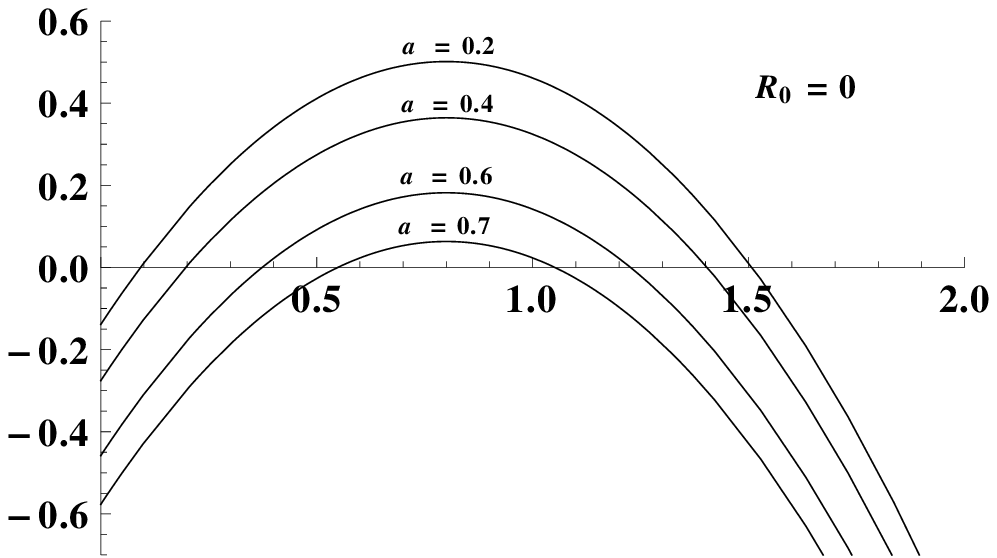}&
\includegraphics[width=9 cm, height=6 cm]{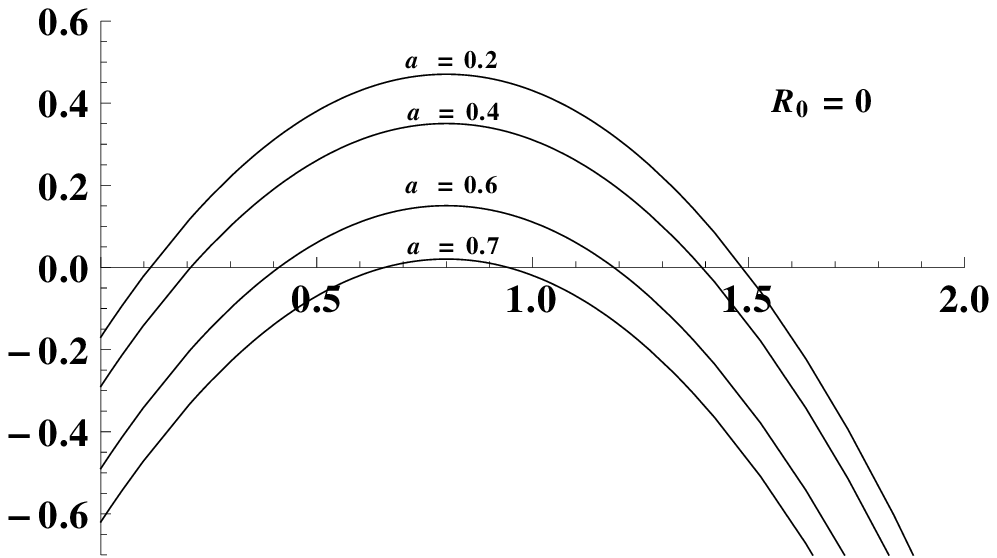}
\\
\hline
\end{tabular}
\caption{\label{AH1} Plots showing the comparison of AH and TLS for
radiating Kerr Newman BH $(R_0=0)$ choosing $M(v)=\lambda v + O(v)$ and $\tilde{Q}(v)=\mu^2 v^2 + O(v^2)$ with parameter values $\lambda=0.04$, $\mu=0.08$ for
different values of rotation parameter $a=0.2,0.4,0.6,0.7$.}
\end{figure}

\end{widetext}

The TLSs of radiating $f(R)$-Kerr-Newman BH is shown in
Fig.~\ref{TLS} for different values of the parameter $R_0$ and rotation
parameter $a$ and surface plot in Fig.~\ref{TLSV} shows the TLSs for
the variable time $v$.  In Fig.~\ref{AH1}, we compared the TLSs and AHs for
different values of rotation parameter of radiating Kerr-Newman BH.  For the definiteness we choose $M(v) = \lambda v +
 O(v)$ and $\tilde{Q}(v) = \mu^2 v^2 +
 O(v^2)$.

\subsection{Apparent Horizon}

The AH is the outermost marginally trapped surface for the outgoing
photons. The AH can be either null or spacelike, that is, it can
`move' causally or acausally \cite{jy}. The AHs are defined as
surfaces such that $\Theta \simeq 0$ \cite{jy}.
Eqs.~(\ref{nvector}) and (\ref{sg}) give the expression for surface
gravity
\begin{equation}\label{sge}
\kappa = \frac{1}{2\Sigma} \left[ \frac{\partial \Delta_r}{\partial
r} - \frac{2r}{\Sigma} \Delta_r\right],
\end{equation}
which on inserting the expression for $\Delta_r$, becomes
\begin{eqnarray}
\kappa &=& \frac{R_0}{12 \Sigma^2} - \left( \frac{R_0}{6 \Sigma }-
 \frac{1- \frac{R_0a^2}{12}}{\Sigma^2}\right)r^3+\frac{2 M(v)}{\Sigma^2}r^2 \nonumber \\
   & & + \left(\frac{1- \frac{R_0a^2}{12}}{\Sigma} - \frac{a^2 +
   \tilde{Q}^2(v)}{\Sigma^2}\right) r - \frac{M(v)}{\Sigma}.
\end{eqnarray}
Eqs.~(\ref{nvector}), (\ref{theta}) and (\ref{sge}) then yield
\begin{eqnarray}\label{thetas}
\Theta &= & - \frac{r}{\Sigma^2} \Delta_r = \frac{r}{\Sigma^2} \Big[
\frac{R_0}{12}r^4 - \left(1- \frac{R_0 a^2}{12} \right)r^2 \nonumber
\\ & & + 2 M(v) r -\left( a^2 + \tilde{Q}^2(v)\right) \Big].
\end{eqnarray}
It is evident that the AHs are zeros of $\Theta=0$.  From
Eq.~(\ref{thetas}), thus the AH's are given by zeros of
\begin{eqnarray}
g(v,r)&=&\frac{R_0}{12}r^4 - \left(1- \frac{R_0 a^2}{12} \right)r^2+ 2
M(v) r
\nonumber \\
&& -\left( a^2 + \tilde{Q}^2(v)\right) = 0. \label{ahe}
\end{eqnarray}

Again in GR limit, we get
\begin{equation}\label{ahgr}
r^2- 2 M(v) r + \left( a^2 + {Q}^2(v)\right) = 0,
\end{equation}
which admit solutions
\begin{eqnarray} \label{ahkn}
  r_{AH}^{-} &=& M (v) -\sqrt {  M^{2}(v)   -
 {a}^{2}    - Q^{2}(v)},
 \nonumber \\
  r_{AH}^{+} &=& M (v) + \sqrt {M^{2}(v)  -
 {a}^{2}    - Q^{2}(v)}.
\end{eqnarray}

\begin{widetext}

\begin{figure}
\begin{tabular}{|c|c|c|}
\hline
\includegraphics[width=9 cm, height=6 cm]{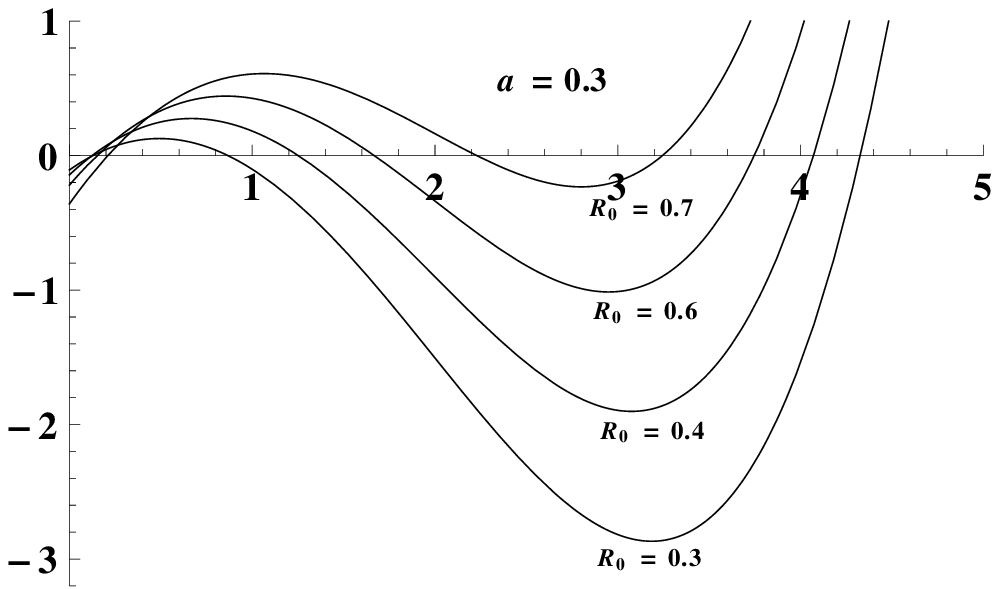}&
\includegraphics[width=9 cm, height=6 cm]{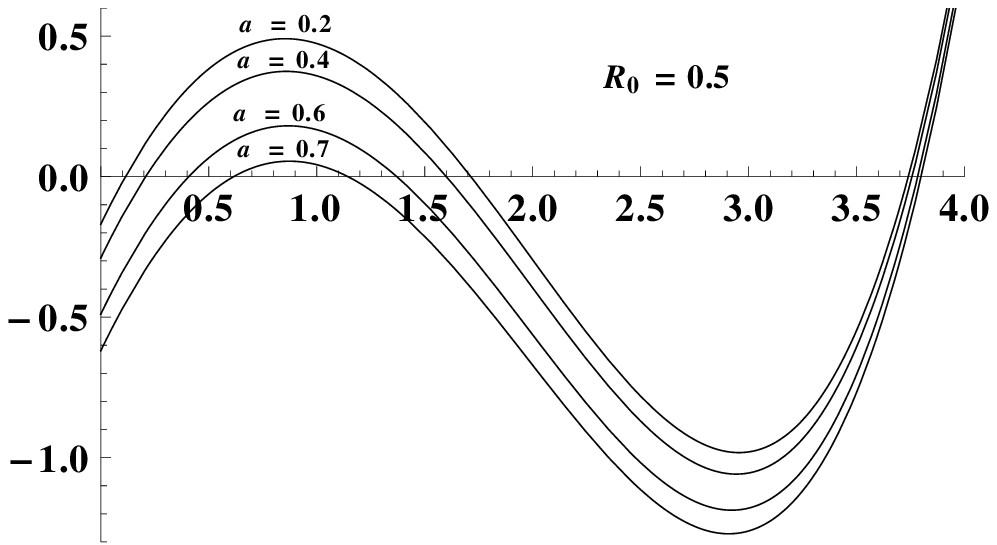}
\\
\hline
\end{tabular}
\caption{\label{AH}AH of radiating $f(R)$ Kerr-Newman BH: Plots showing positive
roots of Eq.~(\ref{ahe}) which corresponds AH of radiating Kerr-Newman BH for $M(v)= \lambda v + O(v)$ and $\tilde{Q}(v)= \mu^2 v^2 + O(v^2)$ with
parameters values $\lambda= 0.04,\; \mu= 0.08$,\; and $\theta=0.3$. The three roots corresponds to $r_{AH}^{-}$ (inner) and
$r_{AH}^{+}$ (outer), and $r_{dAH}$ of radiating $f(R)$ Kerr-Newman BH. LEFT: For different values of $R_0$ with $a=0.3$. RIGHT: For different values of $a$ with $R_0=0.5$. }
\end{figure}

\end{widetext}

\begin{figure}
\includegraphics[width=10cm,angle=0]{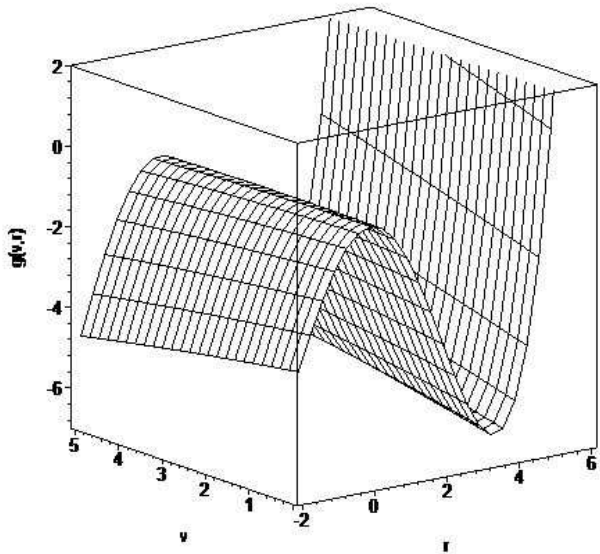}
\caption{A plot of the function $g(v,r)$ for $M(v)= \lambda v +
O(v)$ and $\tilde{Q}(v)= \mu^2 v^2 + O(v^2)$ with parameters values
$\lambda= 0.04,\; \mu= 0.08,\; \theta=0.3$ and $R_0=0.5$}.
\label{AHV}
\end{figure}

There exist, subject to condition \cite{cond}, three positive roots
for $R_0 > 0$ as shown in the Figs.~\ref{AH} and \ref{AHV}.
Unlike, TLS, the AH's are $\theta$ independent. Hence, unlike
non-rotating BHs, they do not coincide in the rotating case.  The
three roots correspond to inner and outer BH AHs, and dS-like AH.
The structure of the AH is depicted in the Fig. \ref{AH}. The surface plot in Fig.
\ref{AHV} shows AH for different values of rotation
parameter $a$ and time $v$ of radiating $f(R)$-Kerr-Newman BH.
For the definiteness we choose $M(v) = \lambda v + O(v)$ and
$\tilde{Q}(v) = \mu^2 v^2 + O(v^2)$.

These are regular outer and inner AHs for a radiating Kerr-Newman
BH, and further  in the non-rotating limit $a \rightarrow 0$,  the
solutions (\ref{ahkn}) correspond to Bonnor-Vaidya AHs. Further,
Eq.~(\ref{ahkn}) in the limit $a \rightarrow 0$ becomes exactly
Eq.~(\ref{tlskn}).  Thus AHs coincide with TLSs, for the non-rotating
but radiating, Bonnor-Vaidya case \cite{sgsd}.

The discussion in above two subsections are also valid for the
stationary case discussed in  \cite{cdr}.  In the stationary
case $M$ and $Q$ are constant whereas in the radiating case $M(v)$
and $\tilde{Q}(v)$ are function of the retarded time $v$.  Thus,
Eqs.~(\ref{tls1}) and (\ref{ahe}) are the same as those derived for the
corresponding stationary case \cite{cdr} when $M(v)=M$ and $\tilde{Q}(v)=Q$
with $M$ and $Q$ constants.

\subsection{Event Horizon}
The EH is a null three-surface which is the locus of outgoing
future-directed null geodesic rays that never manage to reach
arbitrarily large distances from the BH and behave such that $d
\theta / dv \simeq 0$. They are determined via the Raychaudhuri
Eq.~(\ref{re}) to $O(L_M,L_Q)$.  This definition of the EH requires
knowledge of the entire future of the BH.  Hence, it's difficult to find the EH
exactly in non-stationary spacetime.  However, York \cite{jy}, gave
a working definition of the EH, which is in $O(L_M,\; L_Q)$
equivalent to that of photons at EH unaccelerated in the sense that
\begin{equation}\label{EH}
\frac{d^2r}{d n^2}_{|r=r_{EH}} = 0,
\end{equation}
with $d/dn = n^a \nabla_a$.  This criterion enables us to
distinguish the AH and the EH to necessary accuracy. It is known
that \cite{xd99}:
\begin{equation}\label{rdd}
\frac{d^2r}{d n^2} = \frac{1}{\sqrt{A}2\Sigma^2} (r^2 + a^2)
\frac{\partial \Delta_r}{\partial v} + \frac{\Delta_r}{2 \Sigma}
\kappa.
\end{equation}
For low luminosity, the surface gravity $\kappa$ can be evaluated at
AH and the expression for the EH can be obtained to $O(L_M, L_Q)$.
Eqs.~(\ref{rdd}), (\ref{sge}), and the expression for $\Delta_r$,
lead to
\begin{equation}\label{ehe}
\frac{R_0}{12}r^4 - \left(1- \frac{R_0 a^2}{12} \right)r^2+ 2 M^*(v)
r -\left( a^2 + \tilde{Q^*}^2(v)\right) = 0,
\end{equation}
where
\begin{eqnarray*}
M^*(v) &=& M(v) + \frac{(r^2 + a^2)}{\sqrt{A}\; \kappa\; \Sigma} L_M\\
\tilde{Q^*}(v) &=& \tilde{Q}(v)+ \frac{(r^2 + a^2)}{\sqrt{A}\;
\kappa\; \Sigma} L_Q.
\end{eqnarray*}
Eq.~(\ref{ehe}) is the master equation for deciding the EHs of
radiating $f(R)$ Kerr-Newman BH.  It is interesting to note the
mathematical similarity with its counterpart Eq.~(\ref{ahe}) for
AHs. However, unlike the AHs, EHs has $\theta$ dependence as
$M^*(v)$ and $Q^*(v)$ involve $\Sigma$. For stationary BH, $L_M =
L_Q=0$.  Thus, unlike the stationary case \cite{cdr}, where AH=EH
$\neq$ TLS, we have shown that for radiating $f(R)$ Kerr-Newman BH,
AH $\neq$ EH $\neq$ TLS. Thus the expression of the EH is exactly
 the same as its counterpart AH given by Eq.~(\ref{ahe}) with the mass
and charge replaced by the effective mass $M^*(v)$ and charge
$Q^*(v)$ \cite{rm,xd99}.   The GR limit will lead to the same
expression as (\ref{ahkn}), with $M^*(v)$ and $Q^*(v)$ instead of,
respectively, $M(v)$ and $Q(v)$.

\subsection{Ergosphere}
\begin{widetext}

\begin{figure}
\begin{tabular}{|c|c|c|}
\hline
\includegraphics[width=0.30\textwidth]{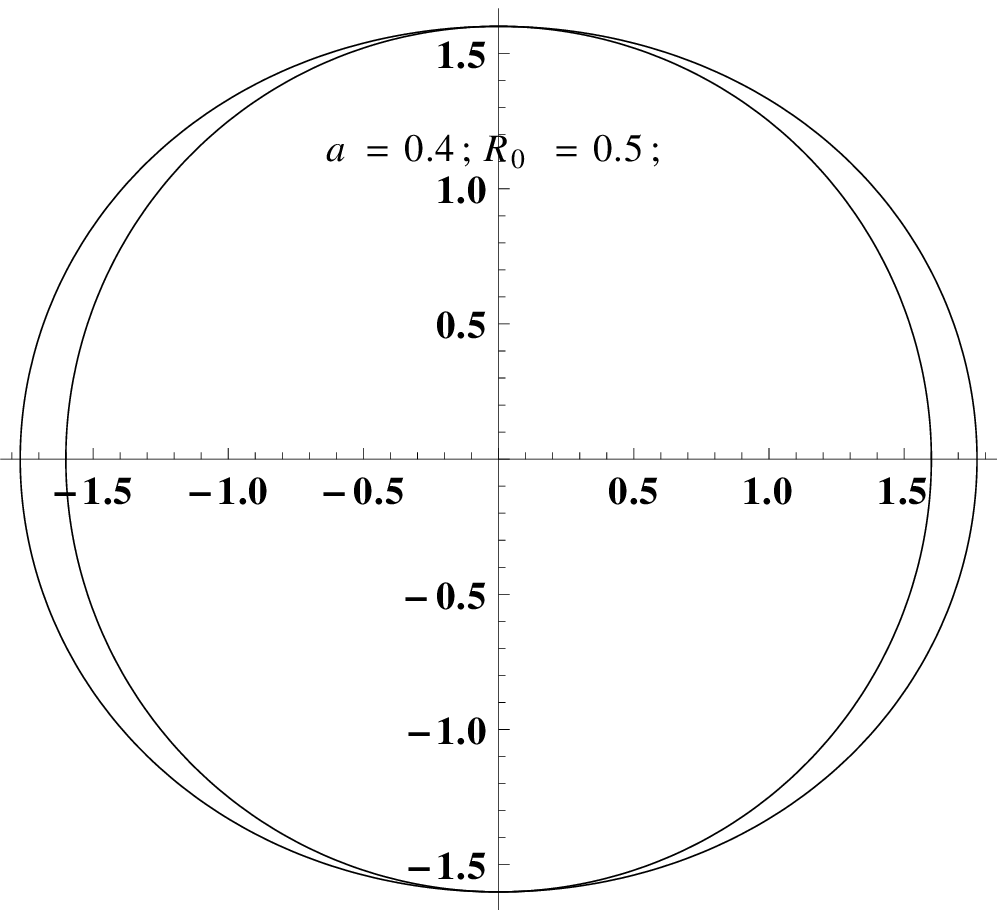}
\includegraphics[width=0.30\textwidth]{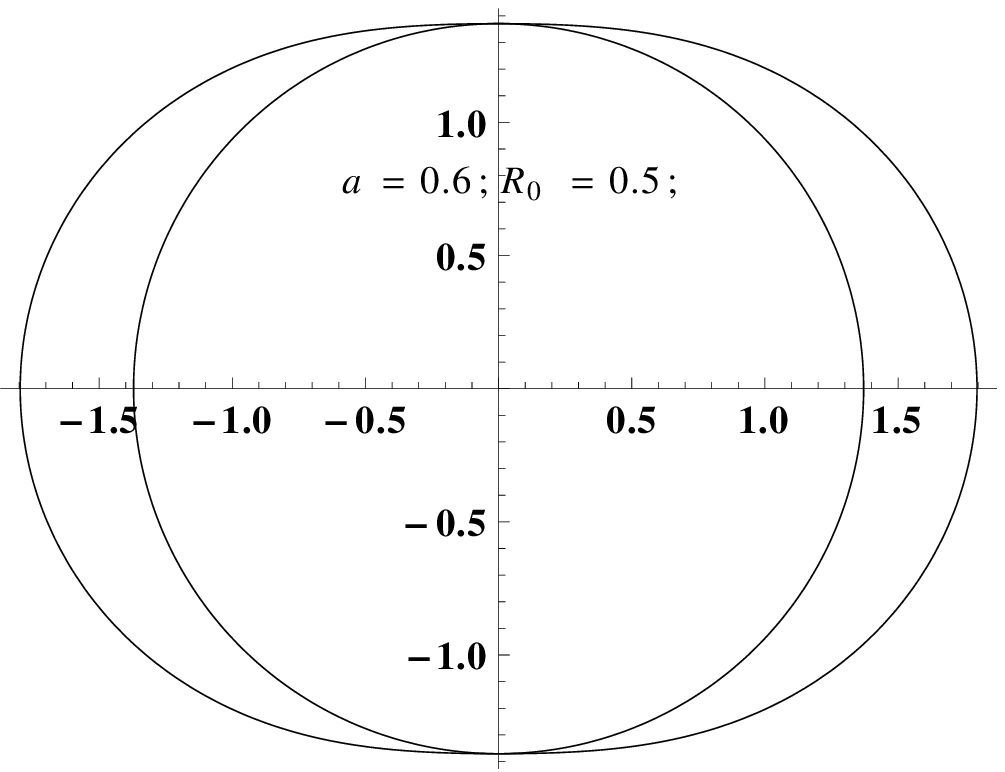}
\includegraphics[width=0.30\textwidth]{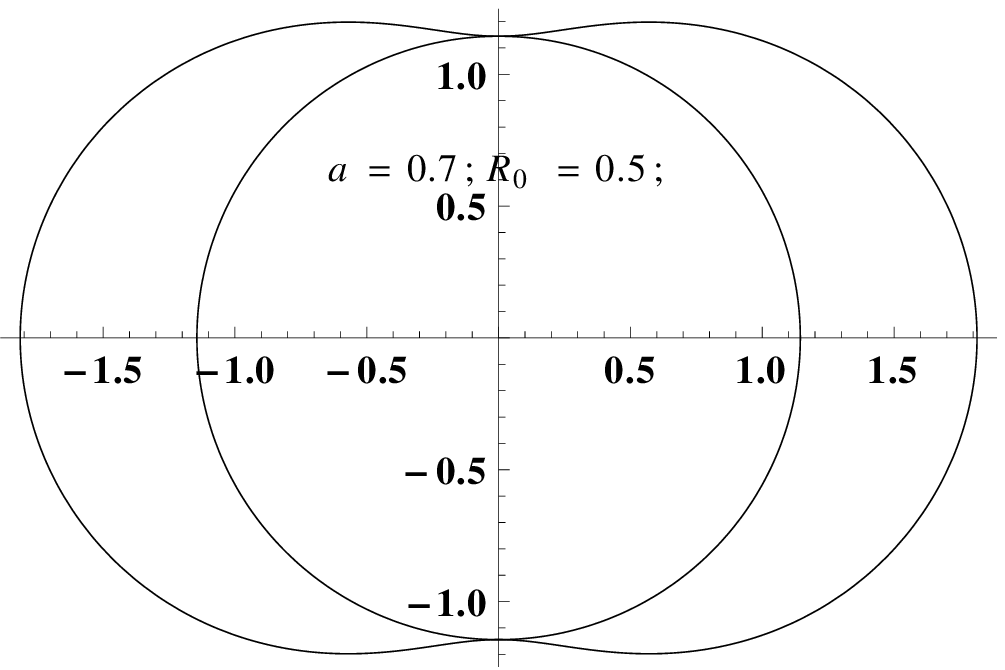}
\\
\hline
\includegraphics[width=0.30\textwidth]{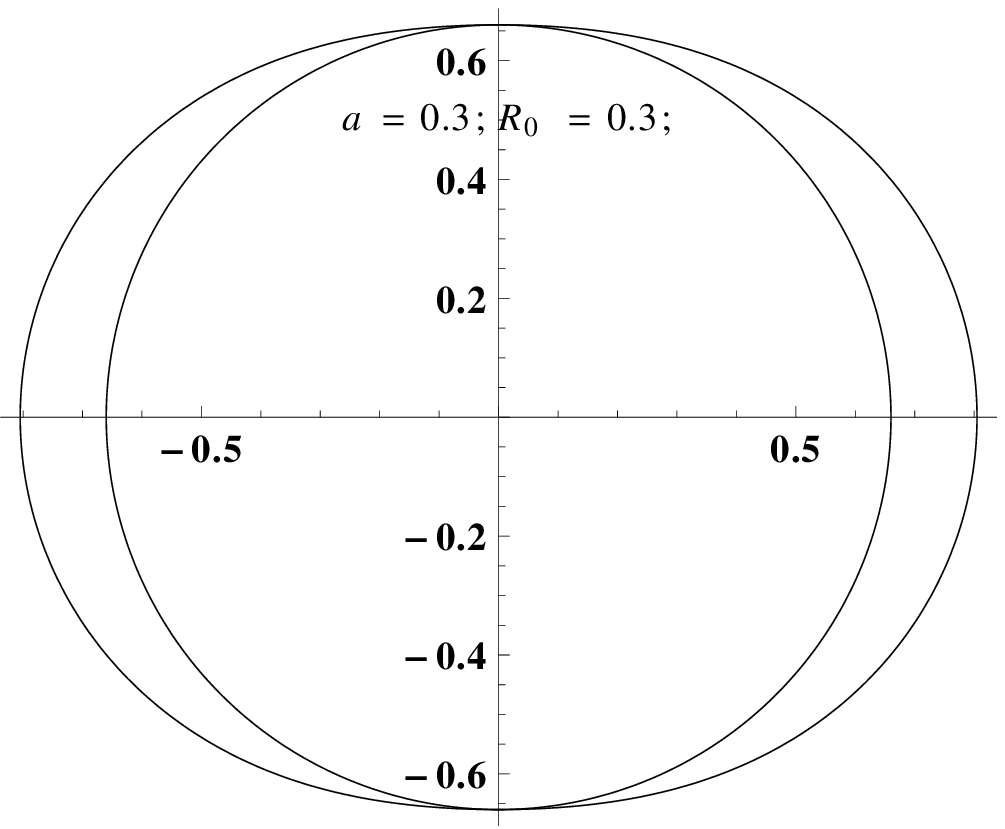}
\includegraphics[width=0.30\textwidth]{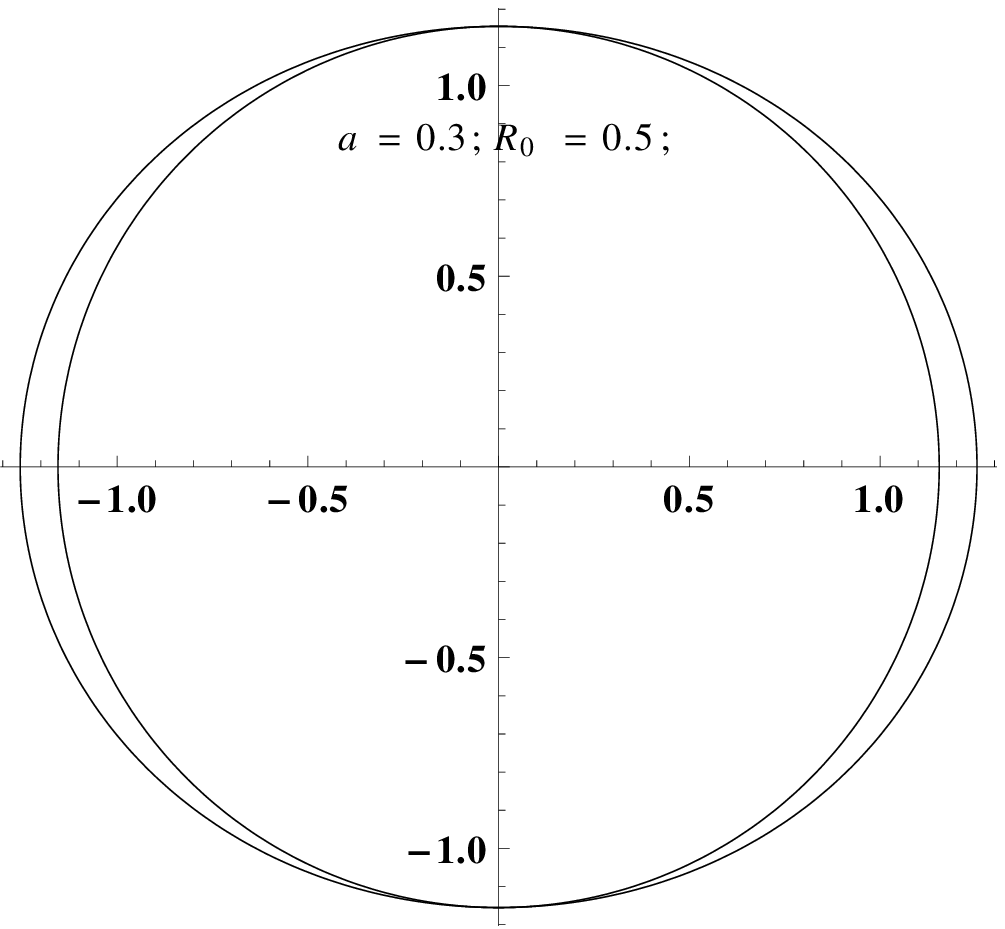}
\includegraphics[width=0.30\textwidth]{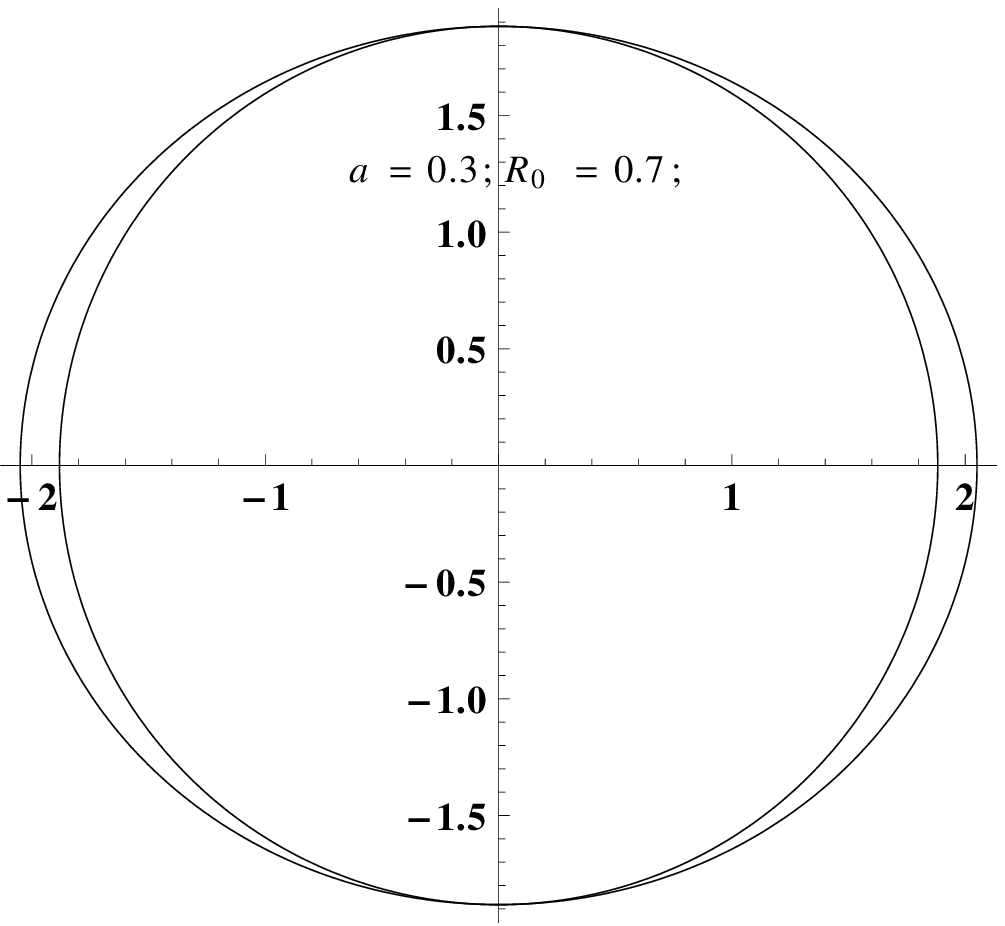}
\\
\hline
\includegraphics[width=0.30\textwidth]{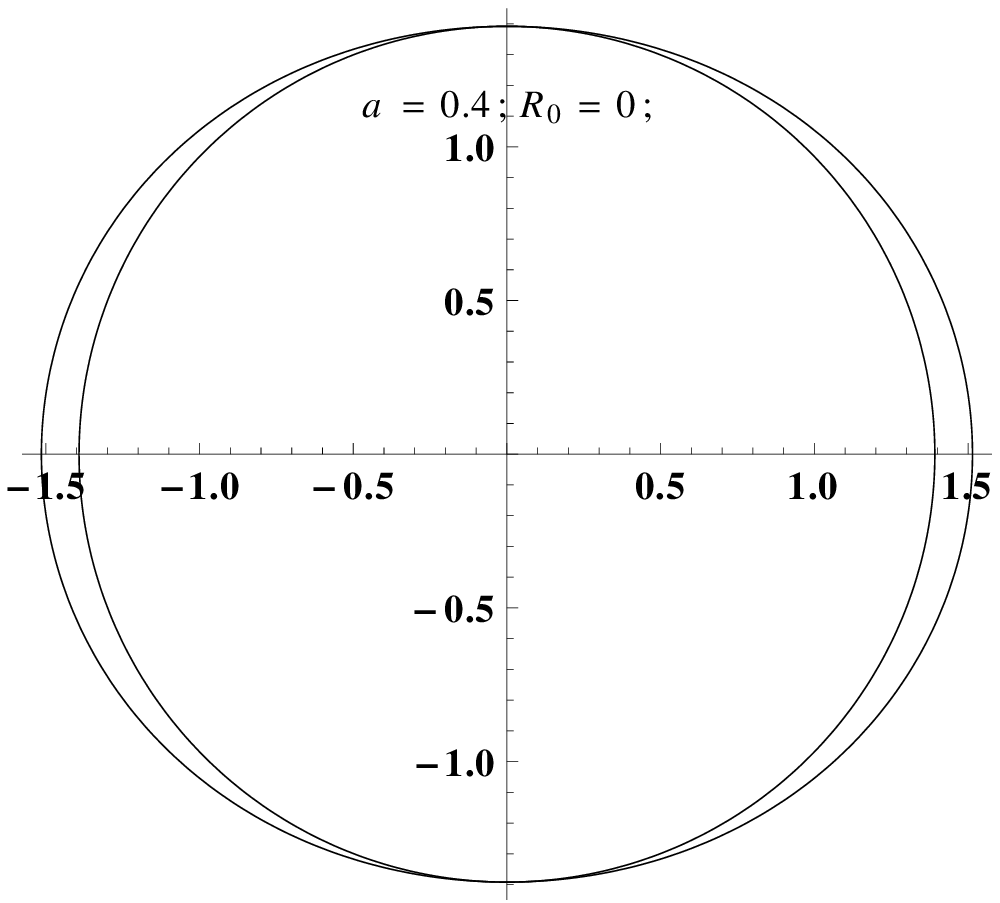}
\includegraphics[width=0.30\textwidth]{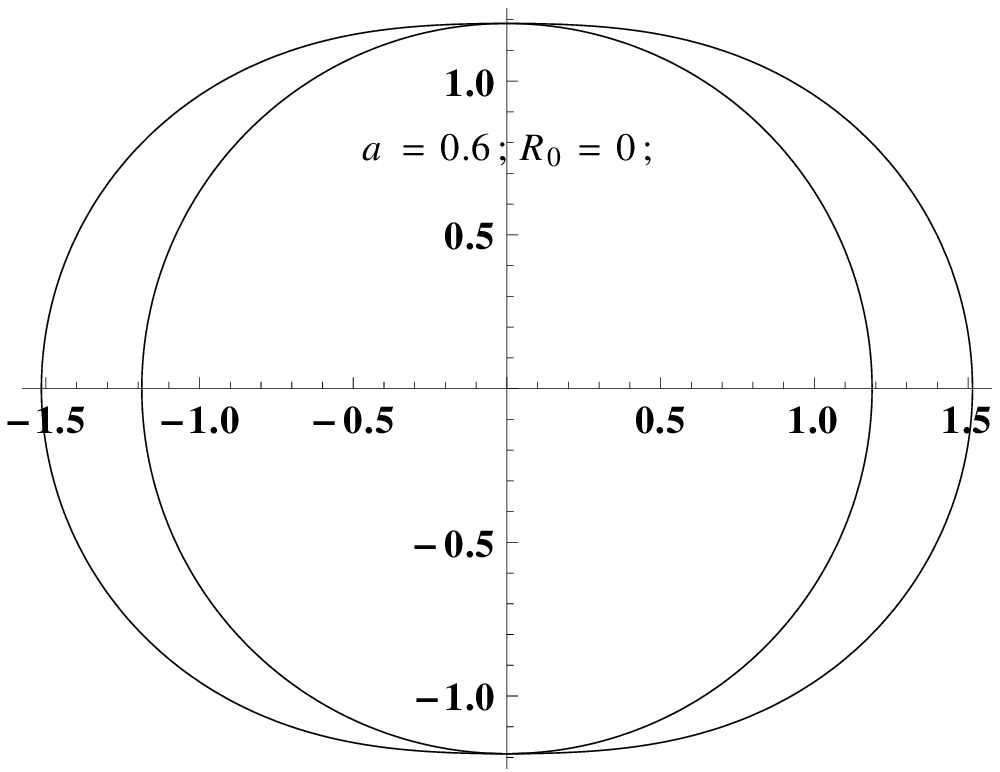}
\includegraphics[width=0.30\textwidth]{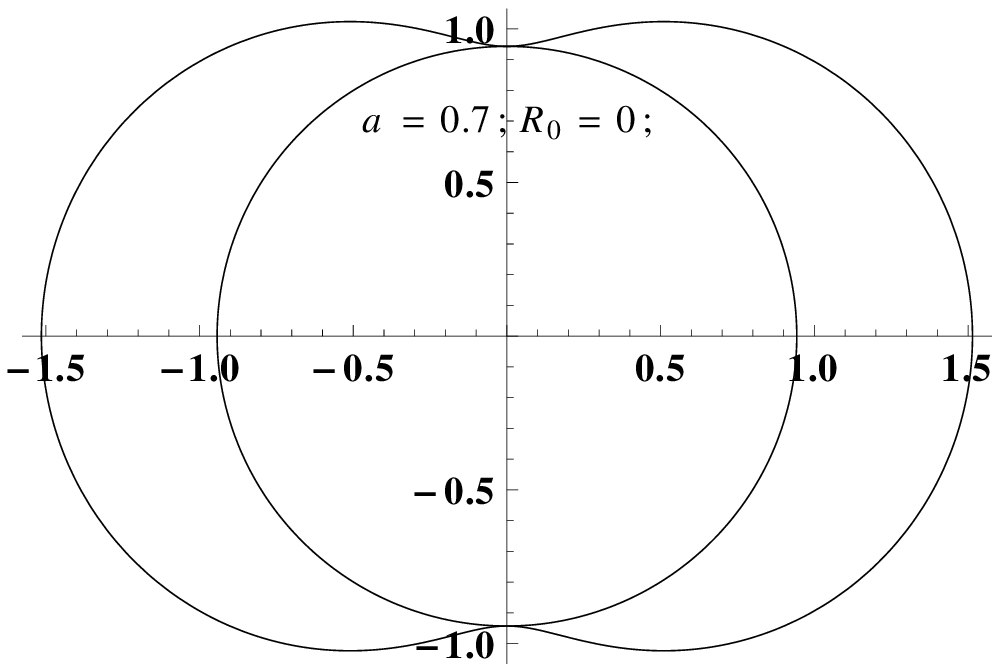}
\\
\hline
\end{tabular}
\caption{\label{ergosphere} Plots showing cross section of horizons
and the shapes of the ergosphere for radiating $f(R)$ Kerr Newman BH
$(R_0>0)$ choosing $M(v)=\lambda v + O(v)$ and $\tilde{Q}(v)=\mu^2 v^2 +
O(v^2)$ with parameter values $\lambda=0.04$ and $\mu=0.08$. UP: For
different values of rotation parameter $a$. MIDDLE: For different
values of $R_0$. DOWN: For different values of $a$ at $R_0=0$.}
\end{figure}

\end{widetext}

For the Schwarzschild and Reissner-Nordstr$\ddot{o}$m BH, it is
possible that a traveler can approach arbitrarily close to the EH
whilst remaining stationary with respect to infinity. This is not
the case for the Kerr/Kerr-Newman BH. The spinning BH drags the
surrounding region of spacetime causing the traveler to spin
regardless of any arbitrarily large thrust that he can provide. The
ergoregion is the region in which this happens and is bounded by the
ergosphere.  The portion of spacetime between horizons and TLS is
called the quantum ergosphere, i.e., the region near the black hole
where negative Killing energies can exist. How the
rotation parameter $a$ and parameter $R_0$ affect the radius of
horizon and shape of the ergosphere for radiating $f(R)$ Kerr-Newman
BH is shown in Figs. \ref{TLS}, \ref{AH} and \ref{ergosphere}.
From the Figs. \ref{TLS} and \ref{AH}, we can see that behaviour
of both the horizon like surfaces of radiating $f(R)$ Kerr-Newman BH
is similar.  The larger the value of parameter $R_0$, the smaller is
the difference between outer  and inner radius of horizons
$(r^+-r^-)$ which affect the size of ergosphere.

By using Eqs. (\ref{tls1}) and (\ref{ahe}), we can draw the
ergosphere of $f(R)$ Kerr-Newman BH. In Fig.
\ref{ergosphere},  we plot the ergosphere for different values of
rotation parameter $a$ and the parameter $R_0$. It is seen that
ergosphere is sensitive to the rotation parameter $a$ as well as the
parameter $R_0$. It is interesting to note that TLS becomes more
prolate thereby increasing the thickness of the ergosphere with increase in
$a$, on the other hand the ergosphere region decreases with the
increase in the value of the parameter $R_0$ which shows that the
ergosphere is sensitive to the parameter  $R_0$.

\section{Conclusion}
 In this paper, we have used a
class of $f(R)$ radiating solutions to generate radiating and
rotating solutions which include the radiating $f(R)$ Kerr-Newman metric
as a special case. The method does not use field equations but works
on the spherical solution to generate rotating solutions. The
algorithm is very useful since it directly allows us to construct
rotating BH, which otherwise could be extremely cumbersome due to
nonlinearity of field equations.
 The $f(R)$ gravity theories are designed to produce a time-varying effective
cosmological constant, the BH and spherically symmetric solutions of
interest are likely to represent central objects embedded in
cosmological backgrounds.  It is evident from the analysis that the
$f(R)$ gravity contributes to a cosmological-like term in the
solutions and they are asymptotically dS/AdS according to the sign
of $R_0$, and has the geometry of the Kerr-Newman dS/AdS.  We have also
established that the Newman-Janis algorithm can be used to derive a
radiating $f(R)$ Kerr-Newman metric.  Originally, the Newman-Janis
algorithm was applied to the Reissner-Nordstr$\ddot{o}$m solution which
is transformed to the Kerr-Newman solution \cite{nja}. The three kinds
of the horizon-like surfaces of the radiating $f(R)$ Kerr-Newman
BHs: TLSs, AHs, and EH were studied by the method developed by York
\cite{jy} to $O(L_M,L_Q)$ by a null-vector decomposition of the
metric. It turns out that  for each of TLS, AH and EH, there exist
three surfaces corresponding to the three positive roots $r^{-}$,
$r^{+}$ and $r_{dH}$. As before $r^{-}$ and $r^{+}$ can be viewed,
respectively,  as inner and outer BH horizons, and $r_{dH}$ as
cosmological or dS-like horizon. The fourth root $r^{- -}$, which is
negative also corresponds to the cosmological horizon \cite{ggsh}.  The
analysis presented is applicable to stationary $f(R)$ Kerr-Newman
BHs as well, but AHs coincide with EHs because stationary BH do not
accrete, i.e., $L_M=L_Q=0$. However,  the three surfaces no more
coincide with each other in radiating $f(R)$ Kerr-Newman BHs. Thus,
we have shown that the presence of the  $f(R)$ gravity term $R_0$
produces a drastic change in the structure of these three horizons.
Such a change could have a significant effect in the dynamical
evolution of these horizons. Thus we have shown that the global
structure of radiating $f(R)$ Kerr-Newman BH is completely different
and far more complicated than that of its GR counterpart.  The
ergosphere is also very sensitive to the term $R_0$ and this in turn
may effect the energy extraction process.

The relation between GR and any modified theory of gravity is a very
good way to know how much of the new theory is different from GR.
Obviously, when $f(R)=R$, the theory reduces to GR. For the
energy momentum tensor (\ref{emt}), the trace $T=0$, consequently
$R$, $f(R)$ and, $f'(R)$  are constant and the theory is equivalent
to GR with a cosmological constant $\Lambda_f=R_0/4$.  Also, the
metric $f(R)$ gravity corresponds to Brans-Dicke (BD) theory with
the potential term, $V(\phi) = f - R f'(R) $, $\phi = 1+f'(R) $ and
the BD parameter $\omega_{BD} = 0$ \cite{frr,lc}.

To conclude, it is notable that there is no exact solution in $f(R)$
gravity coupled to matter with the exception of Maxwell field
\cite{Moon}. We have obtained an exact radiating rotating BH solution in,
constant curvature, $f(R)$ gravity for charged null dust matter. The
solutions presented here provide necessary grounds to
study the geometrical properties, causal structures and thermodynamics
of these BH solutions, which will be subject of a future project.
Further generalization of such solutions  in more general $f(R)$
gravity theories is an important direction \cite{sgg}.

\acknowledgements Two of the authors (S.G.G.) and (U.P.) thank University
Grant Commission (UGC) major research project grant F. NO.
39-459/2010 (SR).  SDM acknowledges that this work is based upon
research supported by the South African Research Chair Initiative of
the Department of Science and Technology and the National Research
Foundation. S.G.G. also thanks Naresh Dadhich for fruitful
discussions.

\appendix

\section{Myers-Perry black holes in $f(R)$ gravity}
Here, we consider a Myers-Perry like solution in constant curvature
$f(R)$ gravity, representing generalization of the exterior metric
\cite{cdm} for the rotating object in $f(R)$ gravity.  However, we
shall restrict ourselves to the uncharged case only because if $D
> 4$, the trace of the electromagnetic EMT is not zero, which is
necessary for finding the constant curvature solutions from $f(R)$
gravity. The HD rotating BH may multiple rotation parameter $a$, but
we shall restricts ourselves to simple case of only one rotation
parameter denoted by $a$.

We have applied the Newman-Janis algorithm to HD $f(R)$ BH metrics
 \cite{cdm} and obtained the corresponding rotating BH metrics given by
\begin{widetext}

\begin{eqnarray} \label{kerrfr}
ds^2&=& \frac{1}{\Sigma}\left[\Delta - \Theta a^2 \sin^2 \theta_1
\right] dt^2- \frac{\Sigma}{\Delta} dr^2 - \frac{\Sigma}{\Theta} d
\theta_1^2
    - \frac{1}{\Sigma}\left[\Theta (r^2 + a^2)^2- \Delta \sin^2\theta_1 \right] \sin^2\theta_1
    d\theta_2^2 \nonumber  \\
    & & - \frac{2 a}{\Sigma}\left[\Theta (r^2 + a^2)-\Delta \right]\sin^2\theta_1 dt d\theta_2 \nonumber -r^2 \cos \theta_1 d
    \Omega^2,
\end{eqnarray}

\end{widetext}
where
\begin{eqnarray*}
d \Omega^2 &= & d\theta_3^2
   +  \sin \theta_3^2 d\theta_4^2  + \sin \theta_3^2  \sin \theta_4^2 d\theta_5^2 +\; \ldots \; \\ \nonumber  & + & \sin \theta_3^2 \times \; \ldots \; \times \sin \theta_n^2
    d\theta_{n+1}^2 \\
\Sigma &= & r^2+a^2\cos^2\theta_1,\\
\Delta &=&(r^2+a^2) \left[1 - \frac{2 R_0 r^2}{(n+1)(n+2)(n+3)} \right]- \frac{2M}{nr^{n-2}}, \\
\Theta &=&1+\frac{2 R_0 a^2}{(n+1)(n+2)(n+3)}\cos^2\theta_1. \\
\end{eqnarray*}
  The solutions are
similar to Myers-Perry dS/AdS solutions. Hence,  we conclude that
the above rotating $D$-dimensional solutions of the constant
curvature $f(R)$ gravity, is just Myers-Perry like solutions in
dS/AdS spacetime  and we call it Myers-Perry solution. For $D=4$,
the metric~(\ref{kerrfr}) reduces to the Kerr metric in $f(R)$
gravity \cite{cdr,al}.  In addition, if $R_0=0$ (GR limit) then
metric~(\ref{kerrfr}) turns out to be the 4D Kerr metric. The
corresponding $D$-dimensional radiating Myers-Perry BH in $f(R)$
gravity can be obtained by  local coordinate transformations
$(t,\;r,\;\theta,\;\phi) \rightarrow (v,\;r,\;\theta,\;\phi)$ and
replacing mass function $M$ by $M(v)$.

\noindent

\begin{thebibliography}{99}
\bibitem{cond} Consider a polynomial of the form $x^4 \;+ \;q\;x^2
\;+\;r\;x+s=0$, with $q,\; r$ and $s$ real and discriminant
$\Delta$, then if $\Delta > 0$ and $s < q^2/4$, then all roots are
distinct and real.
\end{thebibliography}
\end{document}